\documentclass[aps, twocolumn, superscriptaddress]{revtex4} 

\usepackage[switch]{lineno}
\usepackage{amsmath}
\usepackage{empheq}
\usepackage[parfill]{parskip}
\usepackage{graphicx}
\usepackage[active]{srcltx}
\usepackage{color}
\usepackage{array}
\newcolumntype{P}[1]{>{\centering\arraybackslash}p{#1}}
\usepackage{amsmath,amsthm}
\usepackage{dsfont}
\usepackage[T1]{fontenc}
\usepackage[utf8]{inputenc}

\usepackage{graphicx}

\theoremstyle{plain}

\newtheorem*{theorem*}{The curious tourist problem}

\begin{document}

\setlength{\parindent}{0.5cm}

\title{Urban sensing as a random search process}
\author{Kevin O'Keeffe}
\affiliation{Senseable City Lab, Massachusetts Institute of Technology, Cambridge, MA 02139} 

\author{Paolo Santi}
\affiliation{Senseable City Lab, Massachusetts Institute of Technology, Cambridge, MA 02139}
\affiliation{Istituto di Informatica e Telematica, CNR, Pisa, Italy}

\author{Brandon Wang}
\affiliation{Senseable City Lab, Massachusetts Institute of Technology, Cambridge, MA 02139}

\author{Carlo Ratti}
\affiliation{Senseable City Lab, Massachusetts Institute of Technology, Cambridge, MA 02139}

\begin{abstract}
We study a new random search process: the \textit{taxi drive}. The motivation for this process comes from urban sensing in which sensors are mounted on moving vehicles such as taxis, allowing urban environments to be opportunistically monitored. Inspired by the movements of real taxis, the taxi drive is composed of both random and regular parts: passengers are brought to randomly chosen locations via deterministic (i.e. shortest paths) routes. We show through a numerical study that this hybrid motion endows the taxi drive with advantageous spreading properties. In particular, on certain graph topologies it offers reduced cover times compared to random walks and persistent random walks.
\end{abstract}

\maketitle


\section{Introduction}
Random search processes \cite{condamin2007first, benichou2014first, redner2001guide, mendez2014random} are a well studied topic with a bounty of practical applications. Examples include the spreading of diseases and rumours \cite{lloyd2001viruses}, gene transcription \cite{benichou2014first}, animal foraging \cite{viswanathan2011physics,benichou2011intermittent,viswanathan1996levy,bartumeus2005animal}, immune systems chasing pathogens \cite{heuze2013migration}, robotic exploration \cite{vergassola2007infotaxis}, and transport in disordered media \cite{havlin1987diffusion,ben2000diffusion}. Early research on random searches focused on symmetric random walks in Euclidean spaces. Over the years, however, many variants have been considered. These include persistent random walks \cite{tejedor2012optimizing, cenac2018persistent,basnarkov2017persistent} and intermittent random walks \cite{benichou2011intermittent,oshanin2007intermittent, azais2018traveling}, which offer advantages in certain contexts, Lévy flights \cite{shlesinger1986levy, blumen1989transport, viswanathan2000levy, lomholt2008levy} in which the moving particle's jumps are sampled from a heavy tailed distribution leading to non-Gaussian limit laws, and more recently, random walks with memory effects \cite{boyer2014random, schutz2004elephants}. Topologies other than Euclidean spaces, such as random graphs or real-world networks, have also been studied \cite{noh2004random, weng2018universal, masuda2017random, estrada2017random, riascos2014fractional}.

Here, we explore a new random search process: the \textit{taxi drive}. As the name suggests, this process models the movement of taxis. The motivation for studying such a process comes from a recent (theoretical) work in urban sensing \cite{o2019quantifying} in which sensors are deployed on taxis, thereby allowing air pollution, road congestion, and other urban phenomena to be monitored `parasitically'. As such, this drive-by approach \cite{driveby1, driveby2, driveby3, driveby4} to urban sensing can be viewed as a random search process, in which a city's environment is `sensed' (i.e. searched) by sensor-bearing taxis as they drive around serving passengers.

Similar to the run-and-tumble motion of bacteria \cite{schnitzer1993theory,berg2008coli}, the motion of taxis is part-random, part-regular: passenger destinations are chosen randomly, but the routes taken to those destinations are (approximately) deterministic. This mix of regularity and randomness makes the spreading properties of taxis unusual; as shown in \cite{o2019quantifying} -- and reproduced in Figure~\ref{ps_manhattan}(a) -- the stationary distribution of the taxi drive process on real-world street networks follow Zipf's law, in agreement with large, real-world taxi data from nine cities worldwide. This behavior is unusual because it differs significantly from that of classic random search processes such as the random walk, which, as shown in Figure~\ref{ps_manhattan}(c), produces stationary distributions on the same street networks which are skewed and unimodal.

The purpose of this work is to further explore the taxi drive process. We do not study its ability to capture real-world data. Instead, our goal is theoretical: to study the taxi drive as a stochastic process. We focus on cover times, which we numerically compute and compare to those of other well known stochastic processes, namely an ordinary (symmetric, discrete-time) random walk, and the persistent random walk (which we will define later). We hope our paper inspires further work on the taxi drive process and leads to more theoretical interest in urban sensing. \\


\section{Model}

\begin{figure}f
 \includegraphics[width= \linewidth]{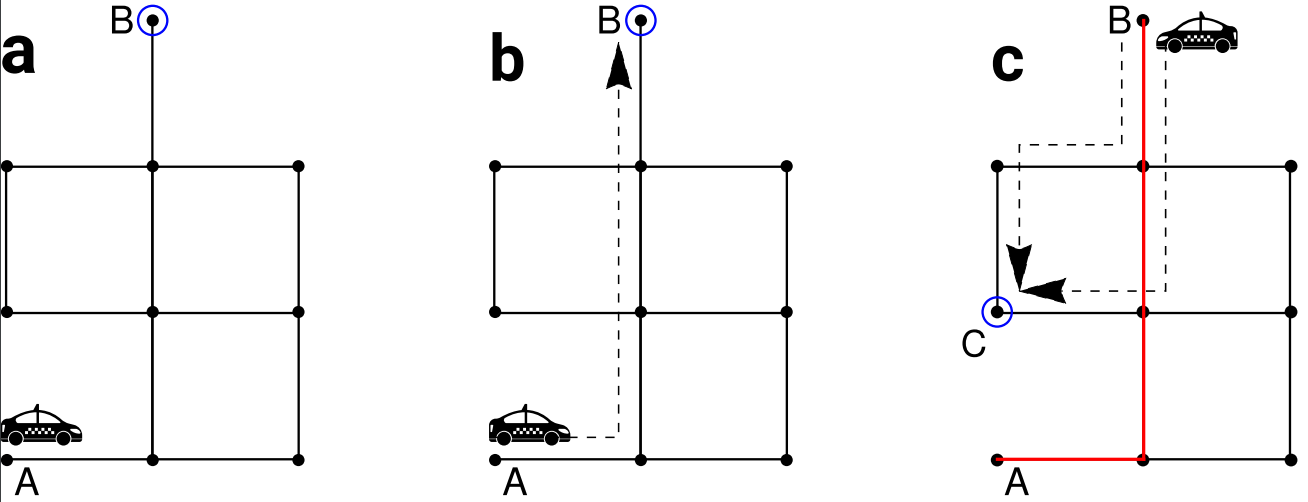}
 \caption{\textbf{The taxi drive}. (a) A taxi picks up a passenger at node A and a destination node B, circled blue, is randomly chosen. (b) The shortest path between $A$ and $B$ is taken as indicated by the dashed arrow. (c) Having dropped off the passenger at $B$, the taxi's next passenger is located at $C$, which is again chosen randomly. There are two shortest paths connecting $B$ and $C$, so one is chosen at random. Having arrived at $C$, the taxi picks up the passengers, and the process repeats.}
 \label{schematic}
\end{figure}

\begin{figure*}
 \includegraphics[width= \linewidth]{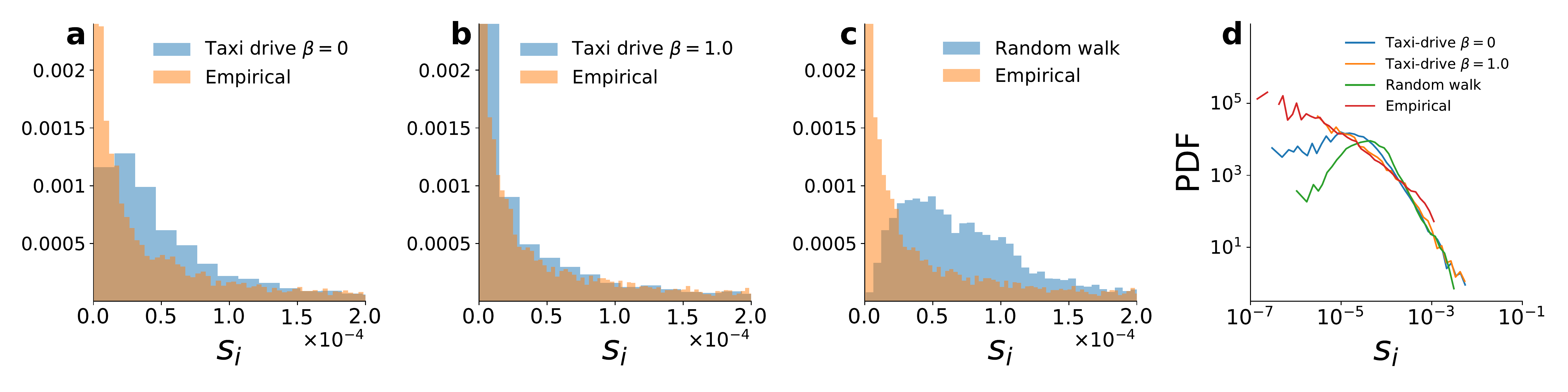}
 \caption{\textbf{Segment popularities}. Histograms of segment popularities $s_i$, defined as the relative number of times the $i$-th street segment (represented by an edge in the street network) is visited on the Manhattan street network. The Manhattan street network was found using the python package `osmnx'. Panel (a): segment popularity distributions from the taxi drive process agree reasonable well with those obtained from empirical data when $\beta = 0$. Panel (b): Better agreement between $s_i^{TD}$ and $s_i^{emp}$ when $\beta = 1$. Panel (c): $s_i$ derived from a random walk do not match the data. Panel (d): Probability density functions of segments popularities. In (a) the taxi drive process was run for $10^7$ time steps and in (b) it was run for $5 \times 10^6$ time steps, after which the $s_i$ distributions were observed to be approximately stationary.}
 \label{ps_manhattan}
\end{figure*}

\textbf{Basic model}. Consider a street network $S$ whose edges represent street segments, and whose nodes represent street intersections. Under the assumption that passengers can only be picked up and dropped off at intersections, nodes represent also possible passenger pickup and dropoff locations. The taxi drive runs on $S$, and as depicted in Figure~\ref{schematic}, is defined by three steps: \\

\begin{enumerate}
    \item A taxi picks up a passenger at node $A$, who wishes to travel to a randomly chosen node $B$.
    \item The taxi travels along the shortest path from node $A$ to node $B$ at unit speed. 
    In the event of multiple shortest paths between $A$ and $B$, one is chosen at random.
    \item Having dropped off the passenger at $B$, the taxi travels to pick up a new passenger at $C$ (again chosen randomly) and the process repeats. \\
\end{enumerate}

We need to specify how the destination node B is chosen. The simplest option is to pick $B$ uniformly at random, which approximately captures the behavior of real taxis. Specifically, it produces reasonably realistic segment popularity distributions, where the $i$-th segment's popularity $s_i$ is the relative number of times that segment was visited by a fleet of taxis over a given reference period (note segments are edges in the street network, and not nodes). Figure~\ref{ps_manhattan}(a) shows the taxi drive segment popularities $s_i^{TD}$ on the Manhattan street network are heavy tailed (Blue histogram; we will explain what the $\beta$ parameter in the legend means soon), in agreement with segment popularities estimated from empirical taxi data $s_i^{emp}$ (orange histogram). Notice however that the  $s_i^{emp}$ are strictly monotonic decreasing, whereas the $s_i^{TD}$ have a small peak; they \textit{increase} over a small interval and then decrease monotonically. In this sense, $s_i^{TD}$ `approximately' capture the behavior of $s_i^{emp}$.

\textbf{Modified model}. The discrepancy between $s_i^{TD}$ and $s_i^{emp}$ can be cured by modifying the model slightly. Instead of choosing destinations uniformly at random, we take inspiration from models of human mobility \cite{song2010modelling} and use a `preferential return mechanism'. Here, the probability of selecting the $n$'th node is $q_n(t) \propto 1 + v_n(t)^{\beta}$, where $v_n(t)$ is the number of times node $n$ has been previously visited up to time $t$ and $\beta$ is a free parameter (Note, the uniformly random choice of destination is recovered in the $\beta = 0$ limit, which explain the legend in Figure~\ref{ps_manhattan}(a)). Figure~\ref{ps_manhattan}(b) shows that $s_i^{TD}$ agrees well with $s_i^{emp}$ when $\beta = 1$. 

The ability of the taxi drive to mimic the behavior of real taxis is not trivial. For example, Figure~\ref{ps_manhattan}(c) shows the $s_i$ resulting from a random walk are skewed and unimodal, in qualitative disagreement with $s_i^{emp}$. Figure~\ref{ps_manhattan}(d) summarizes these findings by showing the probability density functions of the different distributions of $s_i$ plotted in Figures~\ref{ps_manhattan}(a)-(c).

Note that, as described in the Appendix, the dataset the $s_i^{emp}$ was derived from describe the motion of a taxi when it is serving a passenger only; we do not have data on a taxi's movements when it is empty, looking for passengers. Thus when we claim the taxi drive captures the behavior of real-world taxis, we mean specifically the behavior of ``passenger-serving'' taxis only. Whether or not the model also captures the behavior of the passenger-seeking portion of a taxis behavior is unknown.

The ability of the taxi drive to capture the behavior of real taxis was reported in \cite{o2019quantifying}, in which the taxi drive process was originally defined. We include this information here for the convenience of the reader, and to motivate that the taxi drive is model worth studying. But as stated in the Introduction, the goal of this work is theoretical, namely, to study the taxi drive as a stochastic process. With this motivation in mind, we set $\beta = 0$ for the bulk of our work so that destination nodes are chosen uniformly at random. While this means the taxi drive only approximately captures the behavior of real taxis -- as detailed in Figure~\ref{ps_manhattan}(a)  -- it has the benefit of removing the mathematical difficulties imposed by the preferential return mechanism, namely, the spatial memory. Given the discrepancy between $s_i^{TD}$ and $s_i^{emp}$ when $\beta = 0$ is small, we believe this approximation is justified.

\section{Results}

\subsection{Stationary densities}
As a warm up, we consider a well studied property of stochastic processes, namely the stationary density $p_i$, the relative number of times a location $i$ is visited in the large time limit. An elementary result about random walks \footnote{Here we mean the simplest random walk, namely the symmetric, nearest neighbour random walk} on graphs is that $p_i \propto d_i$ where $d_i$ is the degree of node $i$. This is a beautiful result since it exposes the simple connection between the stochastic process (the $p_i$) to the underlying topology the process is run on (the $d_i$). What do the $p_i$ of the taxi drive look like? Do they have a simple relation with the graph topology too? 

\begin{figure}
 \includegraphics[width=0.75\linewidth]{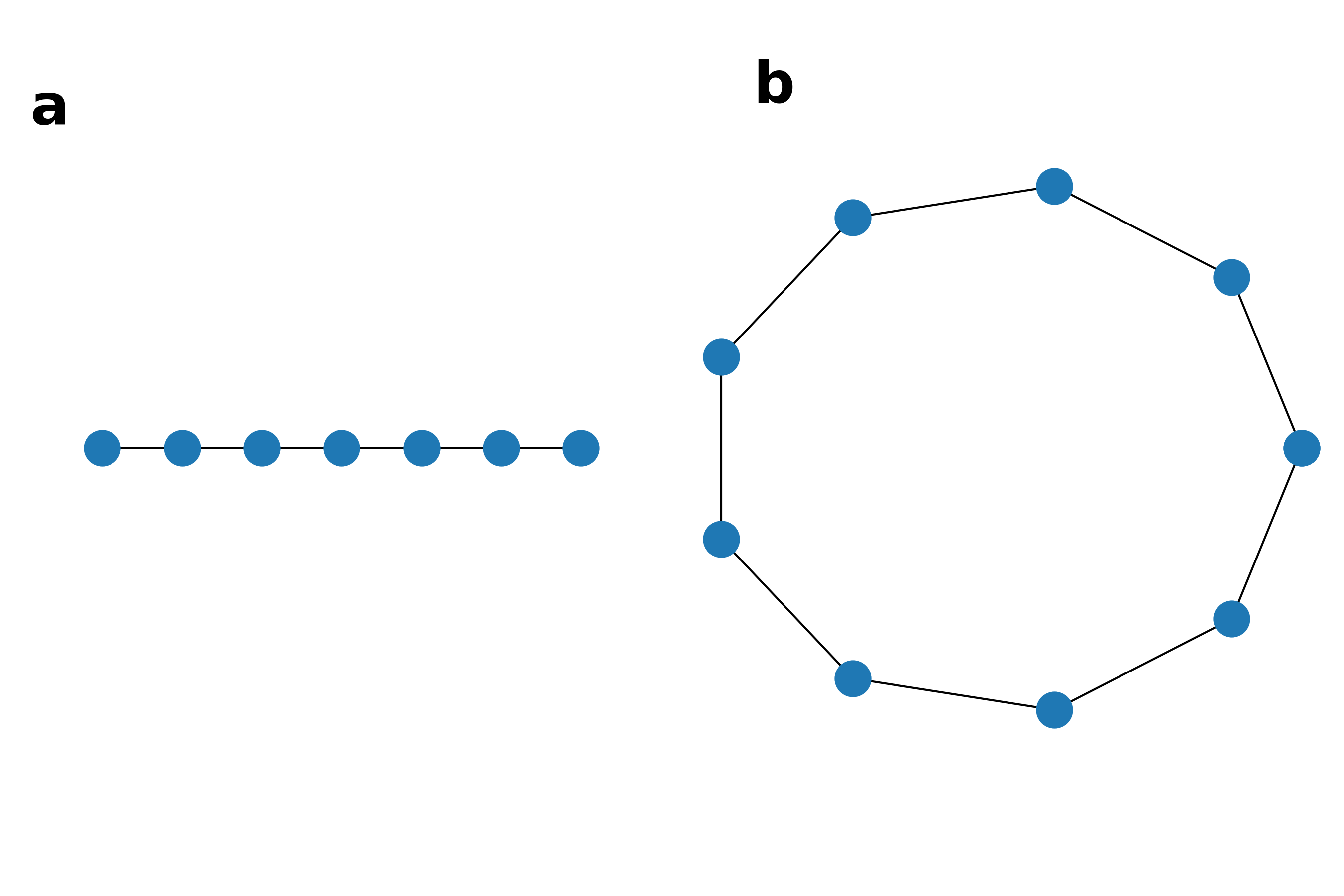}
 \caption{\textbf{Simple graphs} (a) Linear graph (b) Ring graph.}
 \label{graphs} 
\end{figure}

\begin{figure}
 \includegraphics[width=1 \linewidth]{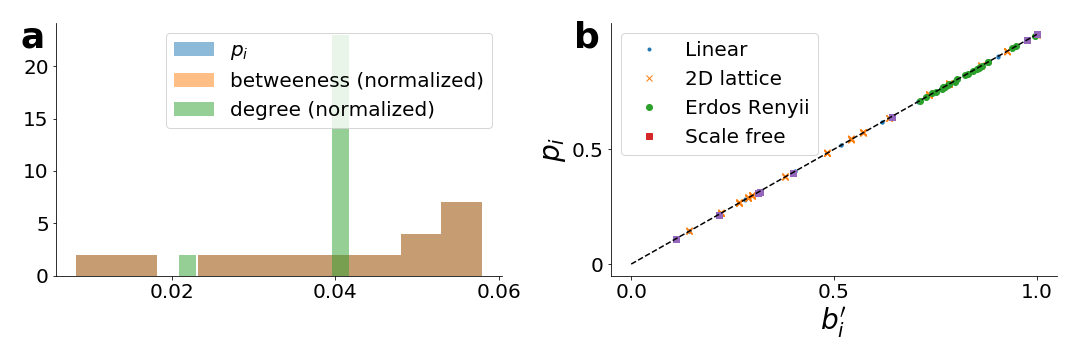}
 \caption{\textbf{Stationary densities} (a) Histogram of the taxi drive density $p_i$, node betweenness $b_i$, and node degrees $d_i$ for the linear graph. We have normalized the betweenness and degree so that their sum is $1$. The taxi drive $p_i$ are distributed similarly to the betweenness (orange and blue shaded areas overlapping perfectly) and are \textit{not} distributed similarly to the degree $d_i$, as would be expected if the $p_i$ were generated from a random walk. The taxi drive $p_i$ were found by running the taxi drive process on the network for $T= 10^5$ timesteps, after which the density appeared stationary. In panel (b) we show $p_i \propto b_i'$  relation for the taxi drive holds true for other graphs, both regular and random. Note $b_i'$ is the adjusted betweenness as defined by \eqref{adj_bet}. Note also that we have re-scaled by $p_i \rightarrow p_i / p_{i,max} $ and $b_i \rightarrow b_i' / b_{i,max}'$ so that the quantities lie in the interval $[0,1]$.}
 \label{stat_dens} 
\end{figure}

We explore this question by studying the taxi drive on the perhaps the simplest graph:  the linear graph which consists of $N$ nodes arranged in a line with edges between adjacent nodes with reflective boundary conditions (Figure~\ref{graphs}(a)). We ran the taxi drive until stable conditions were reached, and counted the relative number $p_i$ of times each node was visited. Note that we focus on nodes here, and not edges as was done in Figure~\ref{ps_manhattan} \footnote{Edges, corresponding to road segments, were the more natural object to consider in the study of urban sensing \cite{o2019quantifying} from which the figure is taken. On the other hand, our focus here is on stationary properties of the underlying graph exploration process, and \textit{node} statistics are typically used for this purpose}. Figure~\ref{stat_dens}(a) shows a histogram of the values of $p_i$ along with histograms of the betweenness $b_i$ (`betweenness' is a measure of the centrality, and is typically defined as the fraction of shortest paths between node pairs that pass through the node of interest; a mathematical definition is given by Eq.~\eqref{x} below) and degree $d_i$ of the nodes in the graph. Interestingly, for the taxi drive the stationary densities of a given node $p_i$ are distributed similarly to the betweenness $b_i$. This contrasts with the stationary densities of the random walk, for which as mentioned, $ p_i \propto d_i$.

The relation between the taxi drive $p_i$ and $b_i$ follows from the similarities between the taxi drive and the definition of a node's betweenness $b_i$, given by
\begin{equation}
    b_i \coloneqq \frac{1}{(N-1)(N-2)} \sum_{s \neq t \neq i} \frac{\sigma_{st}(i)}{\sigma_{st}}
    \label{x}
\end{equation}
\noindent
where $N$ is the number of nodes in the graph, $\sigma_{st}$ is the number of shortest paths connecting the start node $s$ and end node $t$, and $\sigma_{st}(i)$ is the number of those paths that contain the node $i$. Since asymptotically every start-end node combination will be sampled by the taxi drive, and shortest paths are taken between these start and end nodes, one might think $b_i$ and $p_i$ are identical. This is not quite true however. Notice that the start and end nodes are not included in Eq.~\eqref{x}, as denoted by $i \neq s,t$ in the iterator of the sum. This is what makes $b_i \neq p_i$, since in the taxi drive destination nodes are `counted' when they are traversed by the taxi. Origins are however not, since the destination of one trip is the origin of the following trip (taxis spend just one time unit on each node and do not stall when changing direction). Hence a minor modification to Eq.~\eqref{x},
\begin{equation}
    b_i' \coloneqq \frac{1}{(N-1)(N-2)} \sum_{s,t, i \neq s} \frac{\sigma_{st}(i)}{\sigma_{st}},
    \label{adj_bet}
\end{equation}
\noindent
which we call the adjusted betweenness, leads to our final result
\begin{equation}
    b_i' = p_i.
\end{equation}
\noindent
Figure~\ref{stat_dens}(b) shows this relation holds for a variety of different graphs.

In summary, we see that like the random walk, the long term dynamics of the taxi drive (the $p_i$) are related to graph topology (the $b_i'$) in a simple way (which is rare for stochastic processes; the $p_i$ of other common processes such as the Levy walk and persistent random walk do not have clean analytic expressions). Furthermore, it shows that just \textbf{one} graph motif, $b_i'$, influences the long term dynamics; two graphs with different degree distributions $g(d)$ (or any other graph property, for that matter), but identical betweenness distributions $h(b)$ will produce the same $p_i$.


\subsection{Cover times}
We next investigate the cover times of the taxi drive which leads us to pose the following problem: \\ 
\begin{theorem*}
\label{curios}
 A curious tourist arrives in a city with $N$ roads connected in a graph $G$. She decides to explore the city by taking taxis to randomly chosen locations. After being dropped off by a taxi at a given location, she is immediately picked up by another taxi and brought to a new location. How long does it take her to cover every road at least once? \\
\end{theorem*}

\begin{figure}[t!]
  \includegraphics[width=\columnwidth]{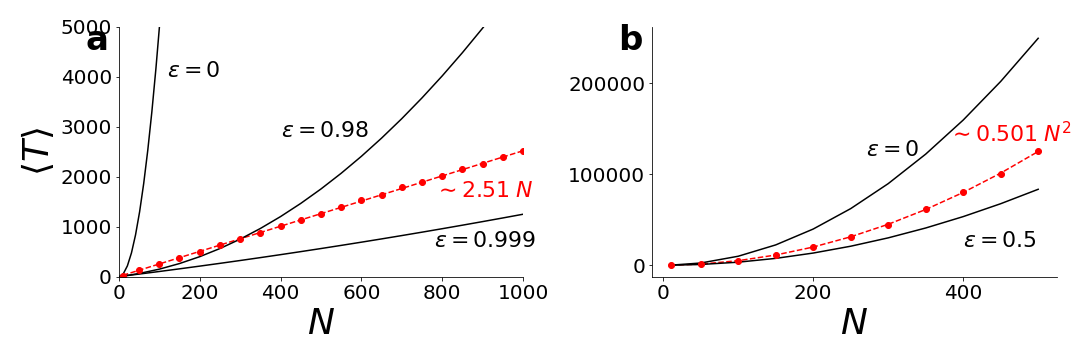}
  \caption{\textbf{Mean cover times on simple graphs}. Red dots show simulation results for the mean cover time of the taxi drive process $\langle T \rangle_{TD}$, while the red dashed line is the lines of best fit to those points. Each data points represents the average of 5000 realizations. Thick black curves show theoretical predictions for the mean cover time of the persistent random walk $\langle T \rangle_{PRW}$ given by Eqs.~\eqref{T_prw1} and \eqref{T_prw2} (the mean cover time of the `ordinary' random walk is recovered at $\epsilon = 0$). Panel (a) Ring graph; for all but extreme values of $\epsilon$ (i.e $\epsilon = 0.999$) $\langle T \rangle_{TD} < \langle T \rangle_{PRW}$. (b) Linear graph; $\langle T \rangle_{TD} > \langle T \rangle_{PRW}$ for modest values of $\epsilon$.} 
  \label{cover_times_ring_path}
\end{figure}

In other words, the curious tourist problem asks to find the cover time of the taxi drive on a graph $G$.

Due to the non-Markovian nature of the taxi drive (the Markovian property is violated since taxis move deterministically when serving passengers; step 2 in definition in the Model Section), we were unable to solve the curious tourist problem analytically (even for the simple, symmetric, nearest neighbour random walk exact results for cover times are rare; see \cite{abdullah2012cover} for a review). So instead we make first attempts at tackling the problem by computing cover times on various graphs numerically.

\textbf{Cover times on simple graphs}. Figure~\ref{cover_times_ring_path} shows how the mean cover time of the taxi drive $\langle T \rangle_{TD}$ varies with graph size $N$ for two simple graphs whose topologies are shown in Figure~\ref{graphs}: the ring graph (a 1D lattice with periodic boundary conditions), and linear graph (1D lattice with reflecting boundary conditions). For comparison, we also plot the mean cover time for the persistent random walk  $\langle T \rangle_{PRW}$. The persistent random walk is a simple extension of the `ordinary' random walk (by `ordinary' we mean the symmetric, nearest-neighbour random walk) with efficient covering properties \cite{tejedor2012optimizing}. Its cover times on the ring and linear graphs are also known exactly, making it a convenient baseline for the taxi drive. It differs from the ordinary random walk in that at each step the walker's direction of motion persists with probability $(1-\epsilon) / 2$ -- and thus a reversal in direction occurs with probability $(1+\epsilon)/2$ --  with $-1 < \epsilon < 1$. The ordinary random walk is recovered at $\epsilon = 0$. Note that in order for the persistent random walker to be well-defined, the graph on which it is run must be embedded in some space, so that a random walker can persist in some `direction'. The ring and linear graphs have this property, and so the persistent random walk can be run on them (for the linear graph, the boundary conditions are reflective, so that the walker changes direction when these endpoints are reached). The reader may be wondering there is some equivalency between the taxi drive process and a random walk with distributed step sizes on the ring and linear graphs; we discuss this in the Appendix.

The cover times for the persistent random walk are \cite{chupeau2014mean}


\begin{align}
     \langle T \rangle_{PRW}^{linear} &= \frac{1-\epsilon}{(1+\epsilon)} (N-1)(N-2) +  \nonumber \frac{2}{(1+\epsilon)}(N-2) \\
     & + \frac{2 + \epsilon}{1 + \epsilon} \label{T_prw1} \\
    \langle T \rangle_{PRW}^{ring} &= \frac{1-\epsilon}{2(1+\epsilon)} N^2 + \frac{5\epsilon - 1}{2(1+\epsilon)}N - \frac{2 \epsilon}{1 + \epsilon}. \label{T_prw2} 
\end{align}
\noindent
The mean cover time of the ordinary random walk $\langle T \rangle_{RW}$ can be found from the above by setting $\epsilon = 0$.

Looking back at Figure~\ref{cover_times_ring_path}, we see the taxi drive covers the linear and ring graphs more efficiently than the ordinary random walk ($\epsilon = 0$) for all graph sizes $N$. The intuition here is that the diffusion constant (or equivalently the persistent length) of the taxi drive is higher than that of the the random walk;  the random walk reverses its direction at every step with probability 0.5 whereas the taxi drive many only reverse its direction when it completes a passenger trip and chooses a new destination. Thus the taxi drive `spreads out' into new terrain (this is roughly what the diffusion constant measures) more quickly than the random walk and therefore reaches every node first. This intuition is however hard to formalize. The problem is the hybrid motion of the taxi drive -- recall when the taxi serving a passenger it moves deterministically, and then changes direction randomly once it has completed this trip -- which does not fit into the markov formalism (which requires homogeneous transition probabilities) which makes analysis difficult. 

How does the persistent random walk fare against the taxi drive? For the ring graph, panel (a), the taxi drive has lower $\langle T \rangle$ for all but extreme values of $\epsilon$ (i.e. $\epsilon \sim 0.999$). For the linear graph, panel (b), however, the persistent random walk beats the taxi drive for moderate values of $\epsilon$ (e.g. $\epsilon = 0.5$). The fact that the persistent random walk beats the taxi drive as $\epsilon \rightarrow 1$ makes sense. This is because at $\epsilon = 1$ the motion is purely ballistic and trivially covers the ring in $N-1$ steps (assuming at $t=0$ the walker has covered the node it starts on, leaving $N-1$ nodes to be covered) and the line in $3/2 N - 1/2$ steps (start at center and walk $N/2$ steps to the left, and then $N$ steps from left to right; the $1/2$ is a corrective factor). 

\begin{figure}[t!]
  \includegraphics[width=\columnwidth]{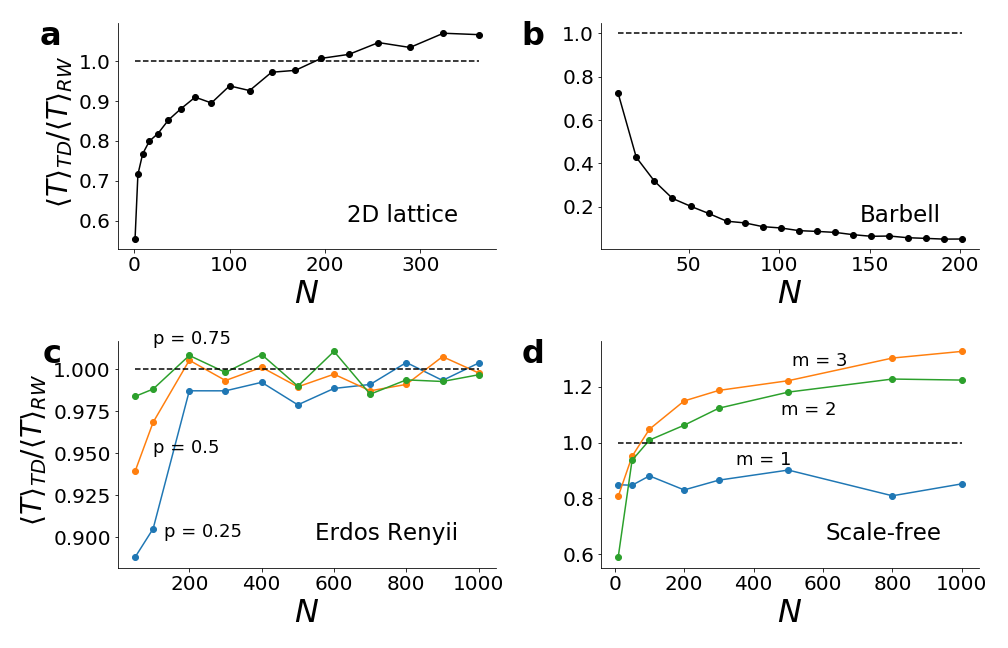}
    \caption{\textbf{Mean cover times on complex graphs}. Dots (both black and colored) represented mean values of ensembles of size $1000$. Thick lines simply connect data points. The dashed black line denotes the value $1$ and is included to highlight the point at which mean cover time for the random walk $\langle T \rangle_{RW}$ exceeds the mean cover time for the taxi drive $\langle T \rangle_{TD}$. For each realization at a given $N$ (there were $1000$ realizations total), a new instance of the random graphs (Erd\H{o}s–R\'enyii and scale-free) was realized which explains the non-monotonicity of the curves. (a) 2D lattice graph. (b) Barbell graph defined in the text (c) Erd\H{o}s–R\'enyii graph with parameter $p$, defined as the probability of there being an edge between any two nodes. This probabilistic generation implies an instance of an Erd\H{o}s–R\'enyii graph could have more than one connected component. In this event, we took the largest of these. (d) Scale-free graph with parameter $m$, defined as the number of edges to attach from a new node to existing nodes during the graph growing process.}
  \label{cover_times_random}
\end{figure}

We next study how $\langle T \rangle_{TD}$ scales as $N \rightarrow \infty$ for the ring and linear graphs. The curves in Figure~\ref{cover_times_ring_path} suggest the ansatz $\langle T \rangle \sim a N^2 + b N$ (the constant term is zero since $\langle T \rangle = 0$ when $N = 0$) so we fit the data to curves of this form. In order for the fitting to be accurate, data at large $N$ must be collected. Beyond $N = O(10^4)$ however simulations were prohibitively costly -- run times being $O(weeks)$ -- so we did not collect data beyond this point. The results of the fitting were  

\begin{align}
     & \langle T \rangle_{TD}^{linear}  = (0.501 \pm 0.001) N^2 + ( -0.49 \pm 2.0) N \label{y1}  \\
    & \langle T \rangle_{TD}^{ring}  = (1.0 \times 10^{-6} \pm 0.1 \times 10^{-7}) N^2 +  (2.52 \pm 0.01) N \label{y2} 
\end{align}
\noindent
From these we conjecture
\begin{align}
    \langle T \rangle_{TD}^{linear} & \sim  O(N^2) \label{y3} \\
    \langle T \rangle_{TD}^{ring} & \sim  O(N) \label{y4} 
\end{align}
\noindent
Eq.~\eqref{y3} follows from Eq.~\eqref{y1}. Eq.~\eqref{y4} follows from Eq.~\eqref{y2}; given the closeness of the coefficient of the $N^2$ term in Eq.~\eqref{y2} to zero we conjecture that $\langle T \rangle_{TD}^{ring} = O(N)$. Of course, since the data these predictions are based on vary over only four decades, we cannot rule out the presence of higher order terms with small coefficients. Thus we restate that Eq.~\eqref{y3} and Eq.~\eqref{y4} are intended as conjectures, whose proofs are open problems.

We remark that the different scaling properties for the ring and linear graphs are puzzling; given the close similarities between the topologies of these graphs, properties of random searches on the graphs (such as cover times \cite{abdullah2012cover}) are typically similar in the $N \rightarrow \infty$ limit.

\textbf{Cover times on complex graphs}. Next we study cover times on graphs with more complex topology, namely 2D lattices, barbells, Erd\H{o}s–R\'enyii, and scale-free graphs. The barbell graph consists of two cliques joined by a single node \footnote{Variants with more than a single node connecting the two cliques have also been considered.}. This topology is known to extremize the spreading properties of the random walk \cite{brightwell1990maximum}, so we include it as a baseline. Figure~\ref{cover_times_random} plots the ratio of $\langle T \rangle_{TD}$ to $\langle T \rangle_{RW}$ -- that is, compares the taxi drive and the ordinary random walk only. We do not study the persistent random walk, since it is ill-defined on graphs not embedded in Euclidean spaces, as the aforementioned graphs are \footnote{The exception being the 2D lattice; persistent random walks have been generalized to two dimensional spaces, but require an additional parameter to specify which direction is chosen when the walk changes its direction of motion. We wished to avoid this complication, so do not study this case.}. Interestingly, as can be seen in Figure~\ref{cover_times_random}, for 2D lattices, Erd\H{o}s–R\'enyii, and scale-free graphs below a certain size, the taxi drive is more efficient than the random walk. This trend appears to hold true for all parameters $p$ of the Erd\H{o}s–R\'enyii graphs, and parameters $m$ of the scale-free graphs (these parameters are defined in the caption of Figure~\ref{cover_times_random}). As expected for the barbell graph, $\langle T \rangle_{TD} < \langle T \rangle_{RW}$ for all graph sizes. The rationale here is that the random walker gets stuck in the bells of the barbell, whereas the ballistic aspect (when the taxi is traveling from origin to destination via the shortest paths) of the taxi drive's movements insulates it from this trapping.


\subsection{m-Cover times}

\begin{figure}
  \includegraphics[width= 0.9 \columnwidth]{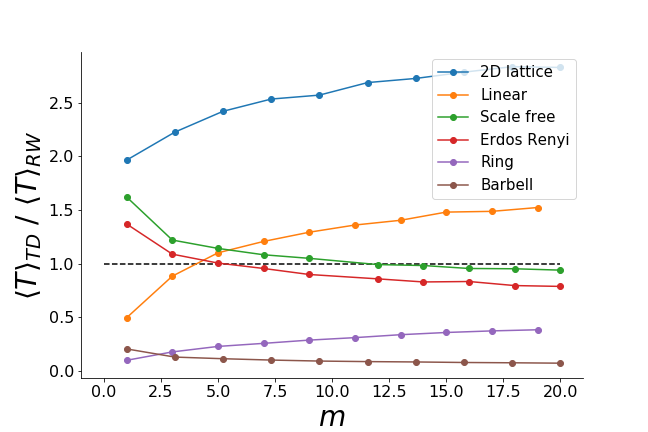}
  \caption{\textbf{$\mathbf{m}$-cover times}. Each colored dot represents the average of $1000$ realizations. Thick lines simple connect data points. The dashed black line denotes 1 and is included to highlight the point at which mean cover time for the random walk $\langle T \rangle_{RW}$ exceeds the mean cover time for the taxi drive $\langle T \rangle_{TD}$. The parameters for each graph were (i) Ring; $N = 50 $ (ii) Linear;  $N = 10 $ (iii) 2D lattice; $N = 400$ (iv) Barbell; $N = 200$ (v) Erd\H{o}s–R\'enyii; $(N,p) = (500, 0.05)$ (vi) Scale-free; $(N,m) = (500, 5)$. The parameters $p$ and $m$ are defined in the caption of Figure~\ref{cover_times_random}.}
  \label{m_cover_times}
\end{figure}

In some random search problems each node needs to be searched more than once. For example in urban sensing, multiple samples at each spatial location are often needed to capture the temporal fluctuations of the quantities being measured, such as air pollution, noise pollution, traffic congestion, and temperature. This leads us to consider the $m$-cover time, the time it takes to cover each node at least $m$ times. Figure~\ref{m_cover_times} plots $\langle T \rangle_{TD} / \langle T \rangle_{RW}$ versus $m$ for the six graphs studied so far. The trends are interesting. For the graphs with regular topology (ring, linear, 2D lattice), $ \langle T \rangle_{TD} / \langle T \rangle_{RW}$ increases with $m$, eventually crossing the threshold value $1$. Yet for the graphs with random topology (Erd\H{o}s–R\'enyii, scale-free) the opposite trend is observed: the taxi drive beats the random walker as $m$ increases. (The barbell graph is an exception here; it has regular topology, but shows a decline in $\langle T \rangle_{TD} / \langle T \rangle_{RW}$ for increasing $m$. This is not too surprising since as discussed its topology maximizes $\langle T \rangle_{RW}$).


\subsection{Preferential return mechanism $\beta$ > 0}
We close by briefly studying how the preferential return mechanism, mathematically given by $\beta > 0$, affects the cover times. The motivation here is to take a step towards  connecting our results to a real world application; as discussed, real taxis are best modeled by the taxi drive process with $\beta > 0 $, so information on cover times in this regime could be useful to practitioners. 

\begin{figure}[h!]
  \includegraphics[width= 1.0 \columnwidth]{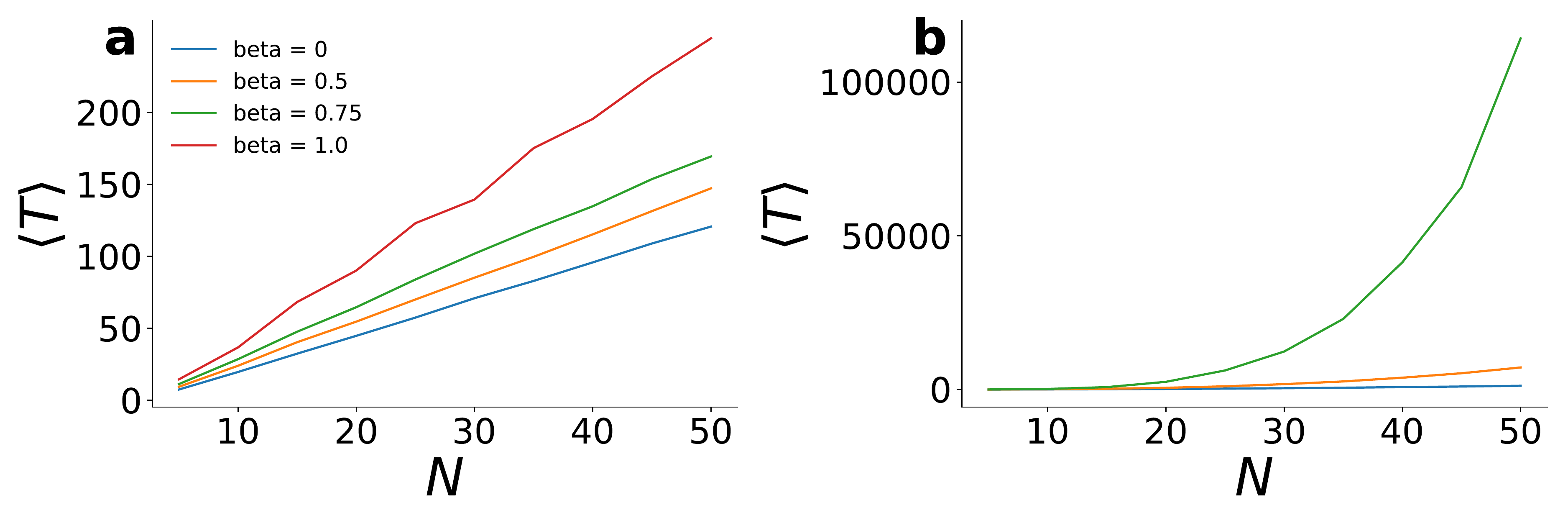}
  \caption{\textbf{Mean cover times with $\beta > 0$}. Panels (a) and (b) shows the mean cover times for the ring graph and linear graphs respectively. For each value of $N$, an ensemble of size $10^4$ was used. Notice graphs of size $ > N = 50$ were not studied (due to prohibitive running times). Notice also that $\beta = 1$ was not computed for the linear graph, again due to large running times.}
  \label{betanonzero}
\end{figure}

We study just the the ring and linear graphs. We expect that $\beta > 0$ will increase the cover times since the probability of selecting an \textit{unvisited} node as a destination goes \textit{down} over time. This is implied by the preferential return mechanism, under which previously visited nodes are preferentially chosen as destination; mathematically, recall, this is expressed by $q_n(t) \propto 1 + v_n(t)^{\beta}$, where $v_n(t)$ is the number of times node $n$ has been previously visited up to time $t$ and $q_n(t)$ is the probability of selecting the $n$'th node (In the Appendix we show $v_n(t) / t \rightarrow const$ so that the taxi drive process is well defined for all $t$). Figure~\ref{betanonzero} shows this intuition is correct: the mean cover times increase with $\beta$ for both the ring and linear graphs.  

The asymptotic scalings of the mean cover times with $N$ when $\beta > 0$ were challenging to estimate. As shown in Figure~\ref{betanonzero}, only graphs up to size $50$ were studied because run times for larger graphs being prohibitively long ($O(weeks)$). Judging by eye, it appears the linear scaling time $\langle T \rangle \sim O(N)$ for the ring graph endures. For the linear graph, however, the quadratic scaling $\langle T \rangle \sim O(N^2)$ seems to break down; fits to polynomials of form $a N^3 + b N^2$ produced parameters $a$ and $b$ of the same order. Of course, given the small range of $N$ the data span over, these fittings are not precise. Analysis and / or more intensive numerics are needed to ascertain the true scaling relations.




\begin{table*}[t]
  \centering
  \begin{tabular}{l | c | l | l |c |c|c}
     & Stationary density $p_i$ & $T_{ring} $ & $T_{line}$ & Graph maximizing $T$ & $T_m$ random graphs & $T_{m}$ regular graphs  \\ \hline \hline
    Random walk (RW) & degree $d_i$ & $O(N^2)$ & $O(N^2)$ & Barbell & -- & -- \\
    Taxi drive (TD) & adj. betweenness $b_i'$ & $O(N)$ (?) & $O(N^2)$ (?) & ??? & -- & -- \\
    Winner & -- & TD  & TD & -- & RW (?) & TD (?) \\
  \end{tabular}
  \caption{\textbf{Summary of results}. $T$ refers to the cover time and $T_{m}$ refers to the $m$ cover time. The adjusted betweenness is defined by \eqref{adj_bet}. A single question mark `?' denotes a numerically backed conjecture. Three question marks `???' denote an open problem. Random graphs refer to Erd\H{o}s–R\'enyii and scale free graphs (because there is randomness in their construction). Regular graphs refer to 1D lattices, 2D lattices and barbells (which do not have randomness in their construction). The persistent random walk values have not been included since the dependence on $\epsilon$ is too hard to summarize in a table. Recall, however, that the persistent random walk is more efficient, as measured by a smaller cover time $T$, than the taxi drive for the ring and line graphs for certain $\epsilon$ as detailed in Figure 4.}
  \label{tab:1}
\end{table*}

\section*{Discussion}

There is a long history of mathematicians taking inspiration from the real world to develop new models and pose new questions. For example, studies of sound vibrations \cite{rayleigh1880xii,rayleigh1896theory} led Lord Rayleigh to introduce the random walk (although Pearson was the first to name it \cite{pearson1905problem}; others \cite{fredriksson2010brief}, however, claim Bernoulli was the first to introduce the random walk when studying the `game of chance' \cite{bernoulli1713ars}). Similarly, the task of routing school buses led Flood to  mathematically pose the traveling salesman problem \cite{lawler1985traveling} \footnote{although others have appeared to study the problem independently \cite{lawler1985traveling}}. In this work, we continued this tradition by taking a cue from urban sensing to introduce the taxi drive. We studied the stationary densities and cover times -- two classic quantities in probability theory -- of this new stochastic process and compared these to the random walk and persistent random walk (natural baselines). In terms of the stationary densities, we found the clean relationship $p_i = b_i'$ which connects the dynamics of the taxi drive to the topology of the underlying graph. In terms of cover times, we posed the curious tourist problem and explored it numerically by introducing conjectures about the scaling relation of $\langle T \rangle $ with graph size $N$. And we also determined the taxi drive outperformed (lower cover time) the random walks in different context. Table~\ref{tab:1} summarizes these findings.

In terms of application, our results show the taxi drive could be useful for random search problems since it outperforms the random walks (typical choices for such problems) in certain contexts. Further, our $p_i = b_i'$ finding could be useful for community detection. Here, efficient algorithms have been designed by exploiting the relationship between the spreading properties of random walkers and graph topology; since the density $p_i(t)$ of a random walker at a given node is related to its degree $d_i$, nodes with similar degree -- that is, nodes which form some `community' -- can be detected by tracking $p_i(t)$ of random walker. By swapping the random walk with the taxi drive, for which $p_i(t)$ is related to $b_i'$, perhaps new types of communities could be cheaply identified. 

In terms of future work, the most pertinent direction would to further explore the curious tourist problem which remains unsolved. For example, our conjectured scalings $\langle T \rangle_{TD}^{ring} = O(N)$ and $\langle T \rangle_{TD}^{linear} = O(N^2)$ ought to be explainable theoretically. More ambitiously, one expects that, given the simpleness of the ring and linear graphs, an \textit{exact} solution (not just scaling relations) for $\langle T \rangle_{TD}$ might be findable. Perhaps the techniques used in \cite{chupeau2014mean} to calculate $\langle T \rangle_{PRW}$, or the techniques used in \cite{maier2017cover} and \cite{chupeau2015cover} which estimate the cover times in terms of the mean first passage time, could be useful for this purpose. Another interesting open problem is to determine which graph topology maximizes the cover time of the taxi drive. The `stickiness' of the bells in the barbell graph traps the random walker -- what counterpart to this graph motif is needed to hamper the taxi drive? Lastly, the behavior of the $m$-cover time could be further analyzed. Why does, as our work suggests, the regularity / randomness of the graph topology determine the scaling of $\langle T \rangle_{TD} / \langle T \rangle_{RW}$ with $m$? We hope future researchers will solve these puzzles.

Source code for the taxi drive is available under the M.I.T. licence and can be found at \cite{code}.

\section*{Acknowledgments}
  The authors would like to thank Allianz,  Amsterdam Institute for Advanced Metropolitan Solutions, Brose, Cisco, Ericsson, Fraunhofer Institute, Liberty Mutual Institute, Kuwait-MIT Center for Natural Resources and the Environment, Shenzhen, Singapore- MIT Alliance for Research and Technology (SMART), UBER, Vitoria State Government, Volkswagen Group America, and all the members of the MIT Senseable City Lab Consortium for supporting this research.


\section{Appendix}

\subsection{Data sets}
The taxi dataset has been obtained from the New York Taxi and Limousine Commission for the year 2011 via a Freedom of Information Act request and is the same dataset as used in previous studies \cite{santi2014}, \cite{tachet2017scaling}. The dataset consists of a set of taxis trips occurring between 12/31/10 and 12/31/11. Each trip $i$ is represented by a GPS coordinate of pickup location $O_i$ and dropoff location $D_i$ (as well as the pickup times and dropoff times). We snap these GPS coordinates to the nearest street segments using OpenStreetMap. We do not however have details on the trajectory of each taxi -- that is, on the intermediary path taken by the taxi when bringing the passenger from $O_i$ to $D_i$. As was done in \cite{santi2014}, we approximate trajectories by generating 24 travel time matrices, one for each hour of the day. An element of the matrix $(i,j)$ contains the travel time from intersection $i$ to intersection $j$. Given these matrices, for a particular starting time of the trip, we pick the right matrix for travel time estimation, and compute the shortest time route between origin and destination; that gives an estimation of the trajectory taken for the trip. Thus, we converted our dataset to a stack of trajectories, where a trajectory is defined by a sequence of road segments. In this format, the street segments popularities $s_i$ are easily derived.

Note that in only having origin $O_i$ and destination $D_i$ GPS coordinates, we do have any information on the taxis movements when it is empty, looking for passengers. We hope future datasets will have this portion of a taxis trips. 

\subsection{Relation between taxi drive and random walk with distributed step sizes on 1D lattices.}

Consider the random walk on the linear graph $G := \{ (1,2), (2,3), (3,4), (4,5) \}$  where $(i,j)$ means nodes $i$ and $j$ share an edge. At each discrete time $t$, the walker draws a step size $L$ from a distribution $p(L)$. Some bounds on the support of $p(L)$ and / or boundary conditions are needed here, but we need not consider them. If the walker is at position $x(t)$ and draws $L = 3$, then the walker travels from $x$ to $x + 3$ at unit speed (as opposed to jumping instantaneously to $x + 3$). In other words the unit speed implies $x(t+1) = x(t) + 1, x(t + 2) = x(t+1) + 1, x(t+3) = x(t+2) + 1$, whereas the instantaneous jump would imply $x(t+1) = x(t) + 3$. 

Notice the step size distribution $p(L)$ does not depend on the position of the walker $x(t)$. This is different from the taxi drive, whose ‘step sizes’ \textit{do} depend on the taxi's position $x(t)$. For example, when at the extremal node $x = 5$,  the taxi’s choices of destination are the nodes $(1,2,3,4)$, corresponding to stepsizes $(-1,-2,-3,-4)$, all chosen with equal probability. Similarly, when at node $x = 1$, the possible stepsizes are $(1,2,3,4)$, again all chosen with equal probability. One could setup the boundary conditions for the random walk so that $p(L)$ would match this step size behavior. But it would be difficult because one would have to reweight the probabilities in $p(L)$ to be uniform over the available options. Thus in this sense, the random walk with distributed step sizes and the taxi drive are different.

\subsection{Taxi drive process with $\beta > 0$}
\begin{figure}[h!]
  \includegraphics[width= 1.0 \columnwidth]{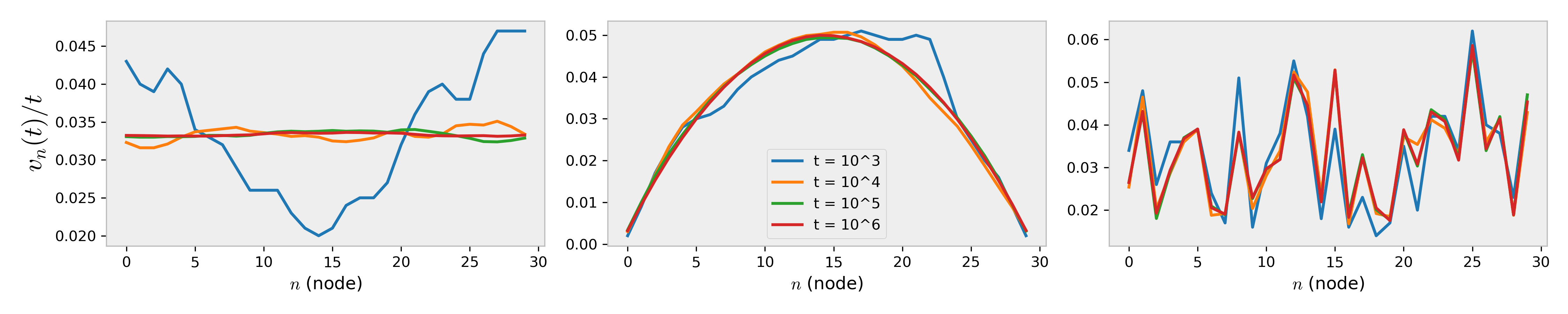}
  \caption{\textbf{Stationary limit $v_{n}(t) / t $}. Each graph has $n = 50$ nodes. In each case a stationary limit is approached. The different curves corresponds to $t = 10^3, 10^4, 10^5, 10^6$ as indicated by the legend in the middle panel. Left: Ring graph. Middle: Linear graph. Right: Erd\H{o}s–R\'enyii graph with $p = 0.2$.}
  \label{stationary-limit}
\end{figure}

When $\beta > 0$ the probability of choosing the $n$'th node as a destination is $q_n(t) \propto 1 + v_n(t)$ where $v_n(t)$ is the number of times node $n$ has been visited up to time $t$. Do the probabilities $q_n$ have a stationary limit? Figure~\ref{stationary-limit} shows they do, as expressed by $v_n(t) / t \rightarrow c_n$ for some constant $c_n$ as $t \rightarrow \infty$. This shows the taxi drive process is well defined as $t \rightarrow \infty$.

\bibliographystyle{apsrev}
\bibliography{ref.bib}

\end{document}